# Kinetic Study of the Reactions of Ground State Atomic Carbon and Oxygen with Nitrogen Dioxide over the 50-296 K Temperature Range

Kevin M. Hickson,[1,*] and Jean-Christophe Loison[1]

[1] Univ. Bordeaux, CNRS, Bordeaux INP, ISM, UMR 5255, F-33400 Talence, France

**Abstract**

The kinetics of the reactions of nitrogen dioxide, $NO_2$, with atomic oxygen and atomic carbon in their ground triplet states ($^3P$) have been studied at room temperature and below using a supersonic flow (Laval nozzle) reactor. $O(^3P)$ and $C(^3P)$ atoms (hereafter O and C respectively) were created in-situ by the pulsed laser photolysis of the precursor molecules $NO_2$ at 355 nm and $CBr_4$ at 266 nm respectively. While the progress of the $O + NO_2$ reaction was followed by detecting O atoms by a chemiluminescent tracer method, progress of the $C + NO_2$ reaction was followed by detecting C atoms directly by vacuum ultra violet laser induced fluorescence at 116 nm. The measured rate constants for the $O + NO_2$ reaction are found to be in excellent agreement with earlier work at higher temperatures and extend the available kinetic data for this process down to 50 K. The present work represents the first kinetics study of the $C + NO_2$ reaction. Although both reactions display rate constants that increase as the temperature falls, a more substantial rate increase is observed for the $O + NO_2$ reaction. The effects of these reactions on the simulated abundances of interstellar $NO_2$ and related compounds were tested using a gas-grain model of the dense interstellar medium, employing expressions for the rate constants of the form, $k(T) = \alpha(T/300)^\beta$, with $\alpha = 1 \times 10^{-11}$ $cm^3$ $s^{-1}$ and $\beta = -0.65$ for the $O + NO_2$ reaction and $\alpha = 2 \times 10^{-10}$ $cm^3$ $s^{-1}$ and $\beta = -0.11$ for the $C + NO_2$ reaction. Although these simulations predict that gas-phase $NO_2$ abundances are low in dense interstellar clouds, $NO_2$ abundances on interstellar dust grains are predicted to reach reasonably high levels, indicating the potential for detection of this species in warmer regions.

# 1 Introduction

Nitrogen oxides such as NO and $NO_2$, collectively known as $NO_x$, are key species in Earth's atmospheric chemistry acting as catalysts in the formation of tropospheric ozone, $O_3$,[1] while in the Earth's stratosphere, $NO_x$ species play a critical role in catalytic $O_3$ destruction cycles.[2] Other nitrogen oxides such as $N_2O$, $NO_3$ and $N_2O_5$ are all known to be present in the Earth's atmosphere. Despite the prevalence of nitrogen – oxygen bonds in terrestrial environments, molecules containing N-O bonds are much rarer in interstellar space. Indeed, among all of the molecules to have been detected in interstellar regions so far,[3] only six of them contain N-O bonds, NO,[4] $N_2O$[5] HNO,[6] HONO,[7] HCNO[8] and $NH_2OH$.[9] Conspicuous by its absence, $NO_2$ has been searched for on several occasions including during a recent survey of the low-mass protobinary IRAS 16293 conducted with ALMA.[7] Here, only a relatively high upper limiting column density of $2 \times 10^{16}$ $cm^{-2}$ could be derived for this species. As a radical molecule (albeit a stable one), the reactivity of $NO_2$ could be relatively high, potentially explaining the lack of detections of this species to date in interstellar space. Indeed, $NO_2$ reacts readily with small molecular radicals such as CH,[10] NH[11] and CN[12] and in particular with all of the atomic radicals H,[11] N,[13] O[11] and C that are considered to be present at high abundance levels in dense interstellar clouds. Nevertheless, at the present time, the C + $NO_2$ reaction has only been investigated at room temperature in a study by Dorthe et al.[14] with no kinetic information given, while the kinetics of the reactions of $NO_2$ with N, O, and H have only been studied at temperatures greater than 200 K[11] due to their potential roles in Earth's atmospheric chemistry. The O + $NO_2$ reaction is particularly important in this respect, as the rate limiting step of the dominant $NO_x$ catalytic ozone destruction cycle in the 25-40 km altitude range.[15] As there is no reliable information regarding the reactivity of $NO_2$ with small radicals at typical interstellar temperatures, it is difficult to constrain simulated $NO_2$ abundances in current astrochemical models. Indeed, by providing a more reliable description of the interstellar chemistry of $NO_2$, it may be possible to target future observations of specific objects where interstellar $NO_2$ abundances are predicted to be the highest.

To address our current lack of knowledge regarding the low temperature reactivity of $NO_2$, we present here a kinetic study of the gas-phase reactions of $NO_2$ with the atomic species C and O in their ground electronic $^3P$ states allowing us to derive rate constants for these processes over the 50-296 K range. The effects of these reactions on interstellar $NO_2$ abundances are tested by including them in an astrochemical model simulating reactivity in both the gas-phase and on interstellar grains and the exchanges that occur between these phases. In order to correctly describe the coupled nitrogen-oxygen chemistry, the present network of reactions has been updated for a more accurate description of the formation and destruction processes for molecules containing N-O bonds and related species. The outline of the paper is described below. The different experimental methods that were applied to follow the kinetics of these reactions are described in section 2, while the results obtained are presented and discussed in the context of previous work in section 3. Section 4 presents the updated astrochemical model and the resulting simulated abundances for molecules containing N-O bonds in the dense interstellar medium. Finally, our conclusions are presented in section 5.

## 2 Experimental Methods

A continuous supersonic flow reactor, also known by the acronym CRESU (Cinétique de Réaction en Ecoulement Supersonique Uniforme or reaction kinetics in a uniform supersonic flow ) was employed to perform the present measurements. The main features of this apparatus were described in detail during the earliest papers using this apparatus. [16,17] Although this system was designed to study the kinetics of reactions between neutral molecular radical species and stable molecules of interest to the chemistry of the interstellar medium and planetary atmospheres via the UV pulsed laser photolysis – UV laser induced fluorescence (LIF) method, [18,19] it has since been modified to extend the range of detection methods to allow the detection of radical atoms such as $C(^3P)$, [20,21] $N(^4S)$ [22] and $H(^2S)$ [23] by vacuum ultraviolet (VUV) LIF in addition to the possibility to detect atomic and molecular radicals by chemiluminescence. [24,25] Three different Laval nozzles were used during this investigation, allowing four different low temperature flows to be produced with Ar and $N_2$ as the carrier gases (one nozzle was used with both gases). The nozzle flow characteristics including properties such as the flow rates and gas densities can be found in Table 1.

**Table 1** Supersonic flow characteristics

| Mach number | 1.83 ± 0.02[a] | 1.99 ± 0.03 | 2.97 ± 0.06 | 3.85 ± 0.05 |
|---|---|---|---|---|
| Carrier gas | $N_2$ | Ar | Ar | Ar |
| Density (×$10^{16}$ cm$^{-3}$) | 9.4 ± 0.2 | 12.6 ± 0.3 | 14.7 ± 0.6 | 25.9 ± 0.9 |
| Impact pressure (Torr) | 8.2 ± 0.1 | 10.5 ± 0.2 | 15.3 ± 0.5 | 29.6 ± 1.0 |
| Stagnation pressure (Torr) | 10.3 | 13.9 | 34.9 | 113 |
| Temperature (K) | 177 ± 2 | 127 ± 2 | 75 ± 2 | 50 ± 1 |
| Mean flow velocity (ms$^{-1}$) | 496 ± 4 | 419 ± 3 | 479 ± 3 | 505 ± 1 |
| Chamber pressure (Torr) | 1.4 | 1.5 | 1.2 | 1.4 |
| Mass flow rate (slpm)[b] | 6.6 | 7.5 | 17.9 | 24.0 |

[a] The errors on the Mach number, density, temperature and mean flow velocity (1σ) are derived from separate impact pressure measurements using a Pitot tube as a function of distance from the Laval nozzle and the stagnation pressure within the reservoir. [b] The mass flow rate is given in standard litres per minutes.

These values were obtained during earlier flow calibration experiments measuring the stagnation pressure within the reservoir in addition to the impact pressure of the supersonic flow via a Pitot tube. For room temperature measurements (296 K), the Laval nozzle was removed from the chamber and the flow velocity was reduced, effectively using the reactor as a slow-flow flash photolysis system.

C and O atoms were generated by different methods in the present study. C atoms were produced by the 266 nm photolysis of trace amounts of the precursor molecule tetrabromomethane, $CBr_4$, entrained in the supersonic flow, with the photolysis laser aligned along the axis of the flow. Pulse energies of approximately 30 mJ were employed here with a 10 Hz repetition rate. $CBr_4$ was carried into the supersonic flow by passing a small flow of the carrier gas into a vessel containing solid $CBr_4$. The gas pressure and temperature of the vessel were fixed for any single series of measurements (pressures between 20 and 120 Torr at 296 K), keeping the gas-phase $CBr_4$ concentration at a constant value. An upper limit of $2 \times 10^{13}$ cm$^{-3}$ was estimated for the $CBr_4$ concentration in the supersonic flow as derived from its saturated vapour pressure at room temperature.

O atoms were produced by the photolysis of $NO_2$ coreagent at 355 nm. This wavelength was preferred due to the weak absorption cross section of $NO_2$ at 266 nm ($\approx 2 \times 10^{-20}$ cm$^2$) compared to the 355 nm value ($\approx 5 \times 10^{-19}$ cm$^2$). At 355 nm the quantum yield for the $NO_2 + h\nu \rightarrow NO + O$ photodissociation process is close to 1.[26] Pulse energies of approximately 50 mJ were employed here with a 10 Hz repetition rate. A 5% mixture of $NO_2$ in Ar was used as the source of $NO_2$ in the present experiments. To verify the gas-phase $NO_2$ concentration, in separate experiments the $NO_2$/Ar mixture was passed through a 10 cm long absorption cell with quartz windows held at room temperature. The 185 nm line of a mercury vapour lamp was used to determine the $NO_2$ absorption in the cell as a function of the cell pressure measured by a capacitance pressure gauge. A value of $2.48 \times 10^{-18}$ cm$^2$ was used for the NO$_2$ absorption cross section at this wavelength following the measurements of Nakayama et al. [27] During the experiments, the NO$_2$/Ar mixture was introduced into the flow upstream of the Laval nozzle reservoir after passing through a calibrated mass flow controller.

Different methods were also employed to follow C and O atoms during the course of the present experiments.

For the experiments on the C + NO$_2$ reaction, C atoms were detected by pulsed VUV LIF at 115.803 nm via the $2s^22p^2$ $^3P_2 \rightarrow 2s^22p5d$ $^3D_3°$ transition. Tunable radiation at this wavelength was produced in a two-step procedure. Firstly, the output of a Nd:YAG laser (532 nm) pumped dye laser around 694.8 nm was frequency doubled in a beta barium borate (BBO) crystal, generating tuneable UV radiation around 347.4 nm. The residual dye beam was discarded by employing two dichroic mirrors coated for 355 nm reflection. The UV beam was then steered towards a cell attached at the level of the observation axis. This beam was focused into the cell using a 20 cm focal length quartz plano-convex lens. The cell contained 50 Torr of xenon while 160 Torr of argon was added for optimal phase matching. The divergent VUV beam generated by this frequency tripling process was then collimated by a MgF$_2$ lens positioned at the exit of the cell, acting as the output window. As the residual UV beam remained divergent due to the large difference in refractive index of MgF$_2$ between UV and VUV wavelengths, a series of circular baffles positioned between the cell and the reactor in the 75 cm long sidearm prevented most of the UV radiation from reaching the supersonic flow while allowing the collimated VUV beam to pass. A supplementary flow of N$_2$ or Ar was added to the sidearm to prevent the residual gases present within the reactor from filling this region and potentially attenuating the VUV beam. The VUV beam crossed the supersonic flow and the photolysis laser (aligned along the supersonic flow) at right angles while emission from unreacted C-atoms in the flow was detected at right angles to the plane containing the photolysis and probe lasers and the supersonic flow, directly above the intersection point of all three using a solar blind photomultiplier tube (PMT). The PMT was not directly attached to the reactor to prevent damage by the reactive gases within the chamber. Instead, a LiF window was used to isolate the PMT from the reactor, with the zone between the window and the PMT held under vacuum by a dry pump to prevent atmospheric absorption of the emitted VUV light. A LiF lens was placed between the LiF window and the PMT within the evacuated region to focus the emitted light onto the PMT photocathode. The PMT output signal was recorded by a boxcar integration system as a function of delay time between lasers with the timing being controlled by a digital delay generator. Each time point consisted of 30 laser shots which were averaged by the control computer with at least 70 time intervals used to establish the kinetic decay profile. Several time points were recorded at negative delays (that is with the probe laser firing prior to the photolysis laser so in the absence of C-atoms), allowing us to record the baseline level for the kinetic traces.

For the experiments on the O + NO$_2$ reaction, O atoms were followed by the chemiluminescence tracer method in a similar manner to the experiments described by Perry.[28] Here, a small flow of NO was added to the flow, leading to the formation of NO$_2$ in an excited state, NO$_2$*, through the termolecular association reaction (R1)

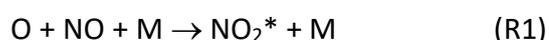

$$O + NO + M \rightarrow NO_2^* + M \qquad (R1)$$

Emission from NO$_2$* was collected at right angles to the supersonic flow and photolysis laser through an interference filter centred at 560 nm (10 nm FWHM) coupled with a long pass filter with a cut-on wavelength at 420 nm to eliminate scattered light from the photolysis laser. A PMT sensitive to radiation in the UV and visible regions was used for this part of the study. The output signal of the PMT was fed into a low noise wideband preamplifier which was connected to a 500 MHz digital oscilloscope for signal acquisition. 1000 time points were recorded for each photolysis laser shot with 768 laser shots being recorded to obtain the final averaged kinetic trace. This technique allowed us to record the entire O-atom kinetic decay for each individual photolysis laser pulse, in contrast to the pulsed laser photolysis pulsed LIF experiments described above where a single time point is recorded for each laser shot. In this respect, the chemiluminescence detection method provides an efficient way to both reduce the time for signal acquisition and improve the signal-to-noise ratio through increased averaging. As reaction (R1) consumes some of the available O atoms, contributing to their overall decay, a fixed concentration of NO was maintained within the flow for any single series of measurements.

To obtain good signal levels, NO concentrations in the range (1-2) × 10$^{15}$ cm$^{-3}$ were typically used for all experiments over the 50-296 K range.

The carrier gas flows Ar and N$_2$ (Messer 99.999 %) and the reactive gases NO$_2$/Ar (Messer 95 % Ar, 5% NO$_2$) and NO (Linde 99.5 %) were regulated by calibrated mass flow controllers. All gases including the Xe (Linde 99.999 %) used in the tripling cell were flowed directly from gas cylinders without further purification.

## 3 Results and Discussion

As $NO_2$ was used in large excess compared to both the minor reagents C and O in the present work, its concentration was assumed to remain constant. Consequently, a pseudo-first-order type analysis of the minor reagent temporal profiles was applied during the analysis of all of the experimental results. In addition, during the experiments to study the kinetics of the O + $NO_2$ reaction, the O + NO (+ M) reaction was used to generate $NO_2$ chemiluminescence. Consequently, the NO concentration was also maintained at a (fixed) excess value for any series of experiments as explained in section 2. It was also necessary to consider the formation of $N_2O_4$ from the $NO_2$ + $NO_2$ + M association reaction in the present experiments Firstly, it was necessary to consider the possible formation of $N_2O_4$ upstream of the Laval nozzle where residence times are long where equilibrium is assumed between $NO_2$ and $N_2O_4$. This was done by considering the formula for the effective second-order rate constant for an association reaction given by Sander et al.,[11] based on the kinetic parameters provided by Atkinson et al.,[29] (including a value for the broadening factor Fc of 0.4). A maximum concentration of $NO_2$ = $1.55 \times 10^{15}$ $cm^{-3}$ was calculated in the reservoir region, due to the large reservoir pressure of 113 Torr when using the 50 K Laval nozzle. In this region, a residence time of 8 ms was estimated for the gas transiting through the reservoir, allowing us to estimate that less than 6 % of the initial $NO_2$ present would dimerize to form $N_2O_4$ upstream of the Laval nozzle. This value was found to vary between 1 and 6 % for the other $NO_2$ concentrations used. Consequently, for the experiments performed at 50 K where the reservoir pressures were high, it is possible that the real $NO_2$ concentrations used are slightly lower than those given by the flow calculations. For the 75 K nozzle, the reservoir pressure is much lower (35 Torr), so that less than 1 % of the initial $NO_2$ was estimated to be in the form of the dimer at the maximum $NO_2$ concentration used in these experiments. Secondly, it was necessary to consider the possible formation of $N_2O_4$ downstream of the Laval nozzle in the cold supersonic flow. Unfortunately, the kinetic parameters provided by Atkinson et al.[29] are only valid over the 300-500 K range, so it was necessary to extrapolate these values down to temperatures beyond their recommended range. Under the most extreme conditions, for our study of the O + $NO_2$ reaction at 50 K, where [Ar] = $2.59 \times 10^{17}$ $cm^{-3}$ and [$NO_2$] = $1.09 \times 10^{14}$ $cm^{-3}$, we estimate that less than 0.5 % of the initial $NO_2$ present formed $N_2O_4$ molecules in the cold supersonic flow (at 50 K, the hydrodynamic time is 350 microseconds).

Although we assume that N$_2$O$_4$ formation is negligible within the supersonic flow itself, it should be noted that this estimate relies on kinetic parameters that are used outside of their validity range. Consequently, further experimental work, such as direct LIF detection of NO$_2$ to study the kinetics of N$_2$O$_4$ formation in the cold supersonic flow should be performed in future to properly characterize the extent of NO$_2$ dimerization in Laval nozzle flows.

**3.1 The C + NO$_2$ reaction**

Representative temporal profiles of the C fluorescence signal recorded at 50 K are shown in Figure 1.

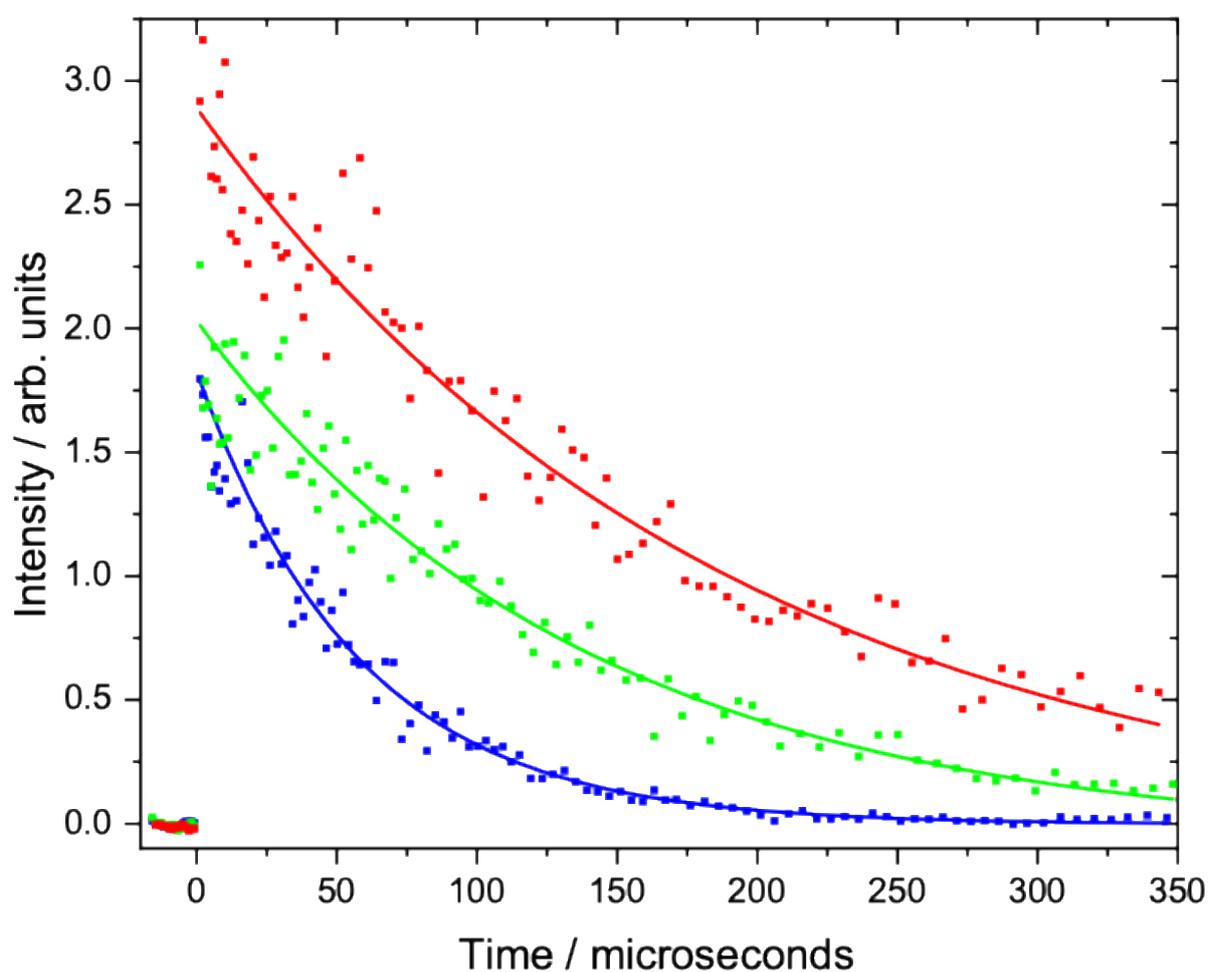

**Figure 1** Atomic carbon VUV LIF signal as a function of decay time recorded at 50 K. (Red squares) without NO$_2$ added ; (green squares) [NO$_2$] = 1.2 × 10$^{13}$ cm$^{-3}$ ; (blue squares) [NO$_2$] = 5.9 × 10$^{13}$ cm$^{-3}$. Solid lines represent non-linear fits to the individual datasets.

It can be seen from Figure 1 that when NO$_2$ is absent from the reactor, as shown by the red datapoints in Figure 1, the C fluorescence signal decays at a non-negligible rate. In this case, C atoms are lost through several secondary processes such as diffusion out of the observation zone (but still within the supersonic flow) with a first-order rate constant $k_{\text{diff}}$ and reaction with CBr$_4$ precursor molecules so that the first-order decay constant, $k_{\text{1st}} = k_{\text{diff}} + k_{\text{C+CBr}_4}[\text{CBr}_4]$. When NO$_2$ is added to the flow, as shown by the green and blue datapoints in Figure 1, the C loss rate increases significantly, indicating a reaction between these two species ($k_{\text{1st}} = k_{\text{diff}} + k_{\text{C+CBr}_4}[\text{CBr}_4] + k_{\text{C+NO}_2}[\text{NO}_2]$). A non-linear least-squares regression method was used to extract $k_{\text{1st}}$ from the individual decay profiles. Kinetic decays were recorded for at least 5 different NO$_2$ concentration values at any single temperature. As $k_{\text{1st}}$ varies as a function of the NO$_2$ concentration, the second-order rate constant, $k_{\text{C+NO}_2}$, could be calculated by plotting the measured $k_{\text{1st}}$ values versus [NO$_2$], as shown in Figure 2 for measurements performed at 50, 177 and 296 K. $k_{\text{C+NO}_2}$ values were finally derived by weighted linear least-squares fits to the data.

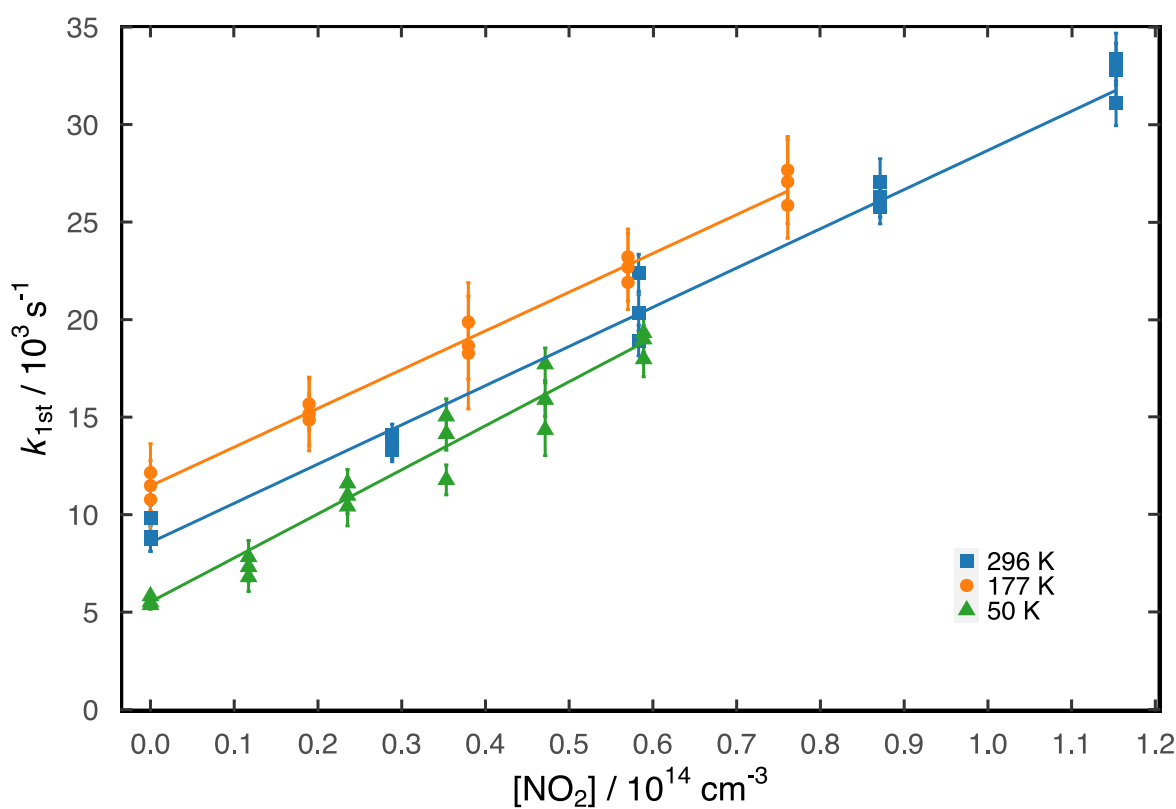

**Figure 2** Variation of the derived pseudo-first-order rate constants with NO$_2$ concentration. (Green solid triangles) data recorded at 50 K in Ar; (orange solid circles) data recorded at 177 K in N$_2$; (blue solid squares) data recorded at 296 K in Ar. Solid lines represent weighted linear least-squares fits to the individual datasets (weighted by the standard deviation attributed to each datapoint, σ, via the expression $1/\sigma^2$). The σ values were derived from exponential fits such as the ones shown in Figure 1.

Following the analysis described above, the y-axis intercept values of the fits to the individual datasets correspond to the constant C-atom secondary loss terms ($k_{\text{diff}}$ + $k_{\text{C+CBr}_4}[\text{CBr}_4]$) for those experiments. The secondary losses of C-atoms are generally lower for those experiments conducted at the lowest temperatures due to the higher dilution factors (larger carrier gas flows with similar flows through the CBr$_4$ vessel) leading to lower CBr$_4$ concentrations. Additionally, flow densities are also higher for these low temperature experiments (see Table 1) so that the contribution of diffusional losses is reduced. Similarly, the flow density used in experiments at 296 K is higher than the one used at 177 K, providing a plausible explanation for the larger measured y-axis intercept value for those experiments conducted at 177 K. The second-order rate constants derived by this analysis and other relevant information are summarized in Table 2. These data are displayed as a function of temperature in Figure 3.

**Table 2** Measured second-order rate constants for the C + NO$_2$ reaction

| T / K | $N^b$ | [NO$_2$] / $10^{13}$ cm$^{-3}$ | $k_{\text{C+NO}_2}$ / $10^{-10}$ cm$^3$ s$^{-1}$ | Carrier gas |
|---|---|---|---|---|
| 296 | 15 | 0 - 11.5 | 2.01 ± 0.21 | Ar |
| 177 ± 2 | 15 | 0 - 7.6 | 2.00 ± 0.21 | N$_2$ |
| 127 ± 2 | 15 | 0 - 8.7 | 2.63 ± 0.27 | Ar |
| 75 ± 2 | 15 | 0 - 4.2 | 2.68 ± 0.28 | Ar |
| 50 ± 1 | 18 | 0 - 5.9 | 2.26 ± 0.24 | Ar |

[a]Uncertainties on the calculated temperatures represent the statistical (1σ) errors obtained from Pitot tube measurements of the impact pressure. [b]Number of individual measurements.

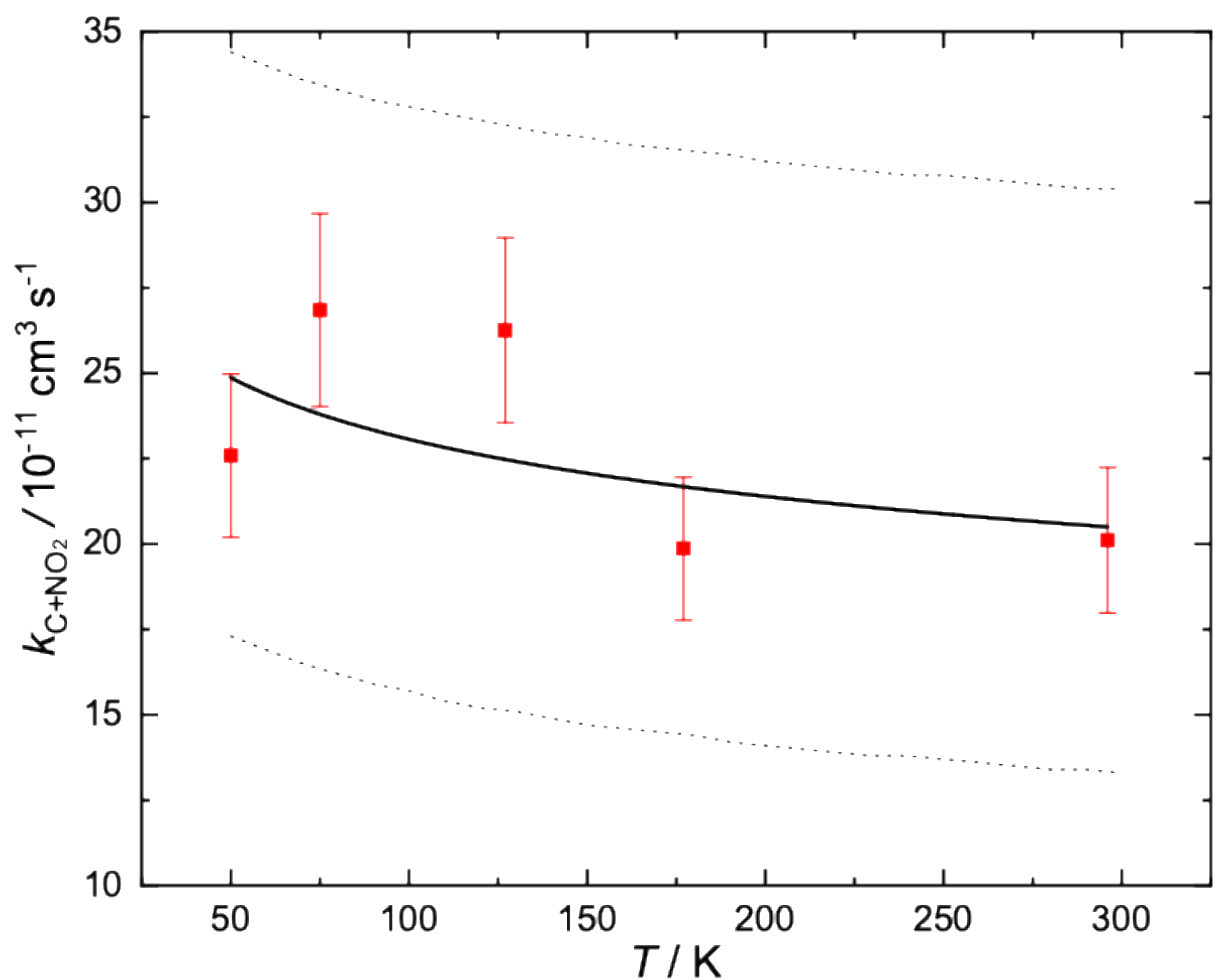

**Figure 3** Temperature dependence of the rate constants for the C + NO₂ reaction. The solid line represents a fit to the data with an expression of the form $k(T) = \alpha(T/300)^\beta$ (see text for more details). The dotted lines represent confidence intervals (30%) on the derived parameters.

It can be seen from Table 2 and Figure 3 that the C + NO₂ reaction is rapid at all temperatures as characterized by rate constants between (2.0-2.7) × $10^{-10}$ cm³ s⁻¹ over the 50-296 K range. The slightly lower measured value at 50 K (compared to those obtained at 75 and 127 K) almost certainly reflects the increased difficulty in the measurements at 50 K due to a lower range of coreagent concentrations used to avoid the possibility of dimer/cluster formation, as well as higher dilution factors leading to lower signal levels. Although there is only a small variation of the rate constant as the temperature falls, the measured values at the lowest temperatures (50, 75 and 127 K) are clearly larger than the ones measured at higher temperature and outside of the combined error bars. In this respect, we have chosen to perform a fit to the experimental data of the form $k(T) = \alpha(T/300)^\beta$ to describe the temperature dependence of the rate constant for this reaction as follows:

$$k(T) = (2.0 \pm 0.9) \times 10^{-10} (T/300)^{(-0.11 \pm 0.09)} \text{ cm}^3 \text{ s}^{-1} \quad (E1)$$

Unsurprisingly, given the small sample size, the rate constants derived from the fit display large confidence intervals. An extrapolation of the fit to 10 K, leads to a rate constant of $k_{C+NO_2}(10 \text{ K}) = 3.0 \times 10^{-10}$ cm$^3$ s$^{-1}$, although this extrapolation is outside of the valid range (50-296 K) for these parameters.

Surprisingly, there is very little information regarding the kinetics of this reaction in the literature, with no earlier measurements of the rate constants for this process. The only previous study of the C + NO$_2$ reaction was performed by Dorthe et al.[14] who investigated the various product channels by observing the wavelength dependent chemiluminescent emission produced. They concluded that the major products were CO + NO with the observed NO formed mostly in the excited B $^2\Pi_r$ state with a weaker contribution from the A $^2\Sigma^+$ state. As this study focused on the chemiluminescent emission from NO, no information was provided about the possible formation of the ground state products NO(X $^2\Pi_r$) + CO(X $^1\Sigma^+$). Interestingly, no chemiluminescent emission was detected from the NO(C $^2\Pi_r$) and NO(D $^2\Sigma^+$) states despite the highly exothermic nature of this reaction (-770 kJ/mol for the formation of NO(X $^2\Pi_r$) + CO(X $^1\Sigma^+$) products, with the NO(C $^2\Pi_r$) + CO(X $^1\Sigma^+$) and NO(D $^2\Sigma^+$) + CO(X $^1\Sigma^+$) products at -143 and -132 kJ/mol respectively). No information was given regarding the possible formation of electronically excited CO products.

## 3.2 The O + NO$_2$ reaction

Representative temporal profiles of the NO$_2$* chemiluminescence signal recorded at 127 K are shown in Figure 4.

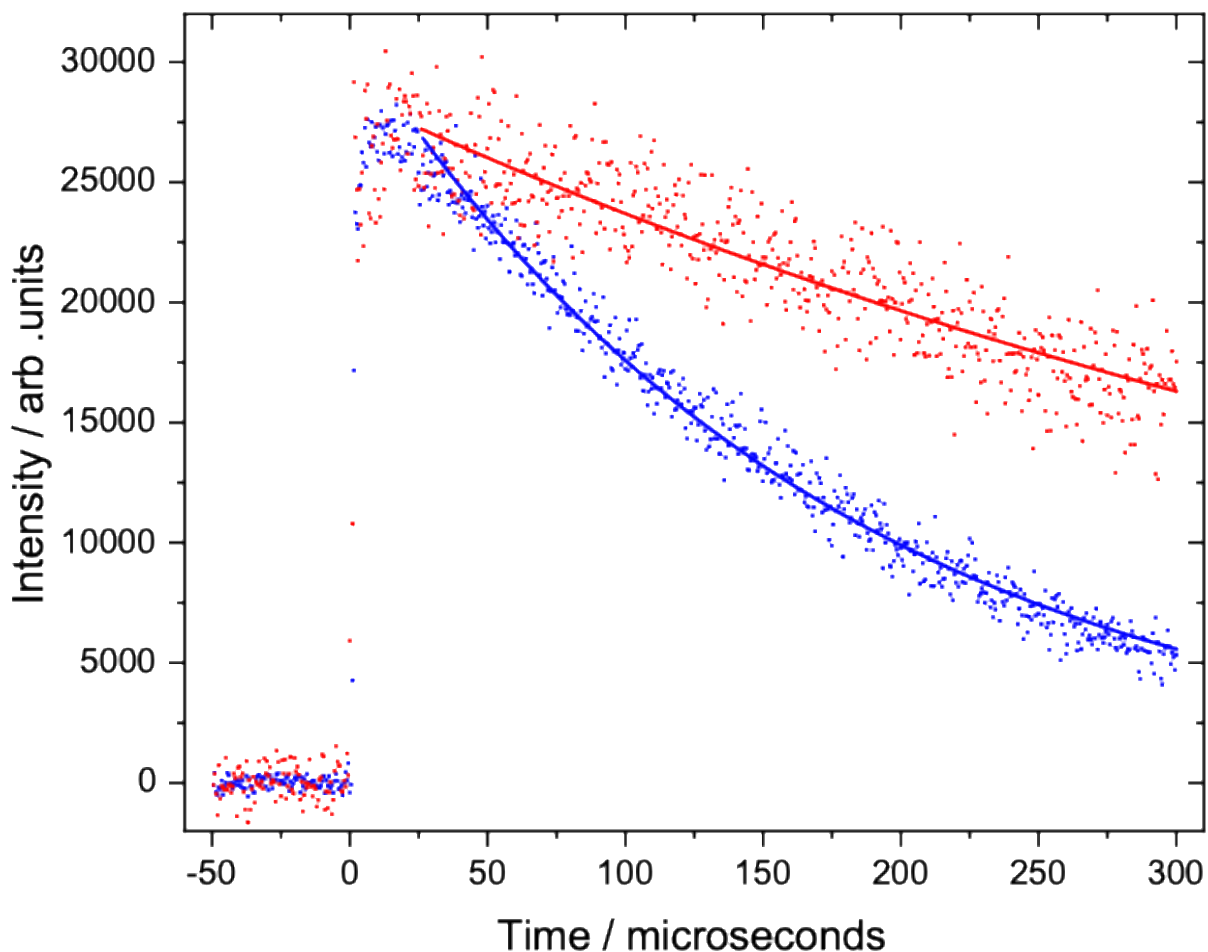

**Figure 4** Temporal decays of the NO$_2$* chemiluminescence signal, tracing the evolution of the O atom concentration, recorded at 127 K. (Red solid points) [NO$_2$] = 6.14 × 10$^{13}$ cm$^{-3}$; (blue solid points) [NO$_2$] = 3.10 × 10$^{14}$ cm$^{-3}$. Solid lines represent the resulting non-linear fits to the individual datasets. The datapoints recorded at low [NO$_2$] have been normalized to the peak value of the datapoints recorded at high [NO$_2$].

There are two main consequences to the use of NO$_2$ as the source of O atoms in the present experiments. Firstly, the intensity of the NO$_2$* chemiluminescence signal used to trace the decay of O atoms was dependent on the NO$_2$ concentration within the supersonic flow. Consequently, those decays recorded at low [NO$_2$] were always characterized by a lower signal-to-noise ratio than the ones recorded at high [NO$_2$]. This is illustrated in Figure 4, where [NO$_2$] = 6.1 × 10$^{13}$ cm$^{-3}$ for the red points, which are noticeably more scattered than the blue points where [NO$_2$] = 3.1 × 10$^{14}$ cm$^{-3}$. It should be noted that the two traces have been normalized to the same peak intensity value to allow a direct visual comparison of the decay rates for these experiments. Secondly, it was not possible to record kinetic decays in the absence of NO₂, lowering the overall range of measured decay rates. Under these conditions, there are three main processes that contribute to the overall loss of O atoms, so that the measured pseudo-first-order rate constant, $k_{1st}$, can be described by expression (E2),

$$k_{1st} = k_{diff} + k_{O+NO}[NO][M] + k_{O+NO_2}[NO_2] + k_{O+NO_2+M}[NO_2][M] \qquad (E2)$$

where $k_{diff}$ is first-order diffusional loss of O atoms from the observation region, $k_{O+NO}[NO][M]$ is a term representing the loss of O atoms through the termolecular tracer reaction with NO molecules, $k_{O+NO_2}[NO_2]$ is the term representing the loss of O atoms through the bimolecular target reaction leading to O₂ + NO as products and $k_{O+NO_2+M}[NO_2][M]$ represents O atom loss through the O + NO₂ + M termolecular association reaction leading to NO₃ + M as products. As we follow the loss of O atoms, it was not possible to discriminate between the bimolecular and termolecular channels of the O + NO₂ reaction. Nevertheless, based on the temperature dependent parameters provided by Sander et al.[11] for the O + NO₂ association reaction, we estimate that this process contributed at most 6 % to the total rate constant (at 50 K) with negligibly small contributions at all other temperatures. As the O + NO + M reaction might represent a significant loss process for O atoms, it was necessary to estimate its contribution to the overall decay rate. If we use the temperature dependent rate parameters given by Sander et al.[11] for this termolecular association (valid over the 200-300 K range) and extrapolate them to lower temperature, multiplied by the relevant flow density (see Table 1) and NO concentration (in the range (0.75-2.09) × 10¹⁵ cm⁻³), we obtain negligibly small pseudo-first-order loss rates of less than 100 s⁻¹ over the 75-296 K temperature range with a slightly larger estimated value of 360 s⁻¹ at 50 K. As the terms $k_{diff} + k_{O+NO}[NO][M]$ make a constant contribution to the individual decay curves for any series of measurements at a single temperature, these are reflected in the y-axis intercept values of the second-order plots shown in Figure 5. As these values fall in the range 600-1200 s⁻¹, the major secondary losses of O atoms are likely to be due to diffusion.

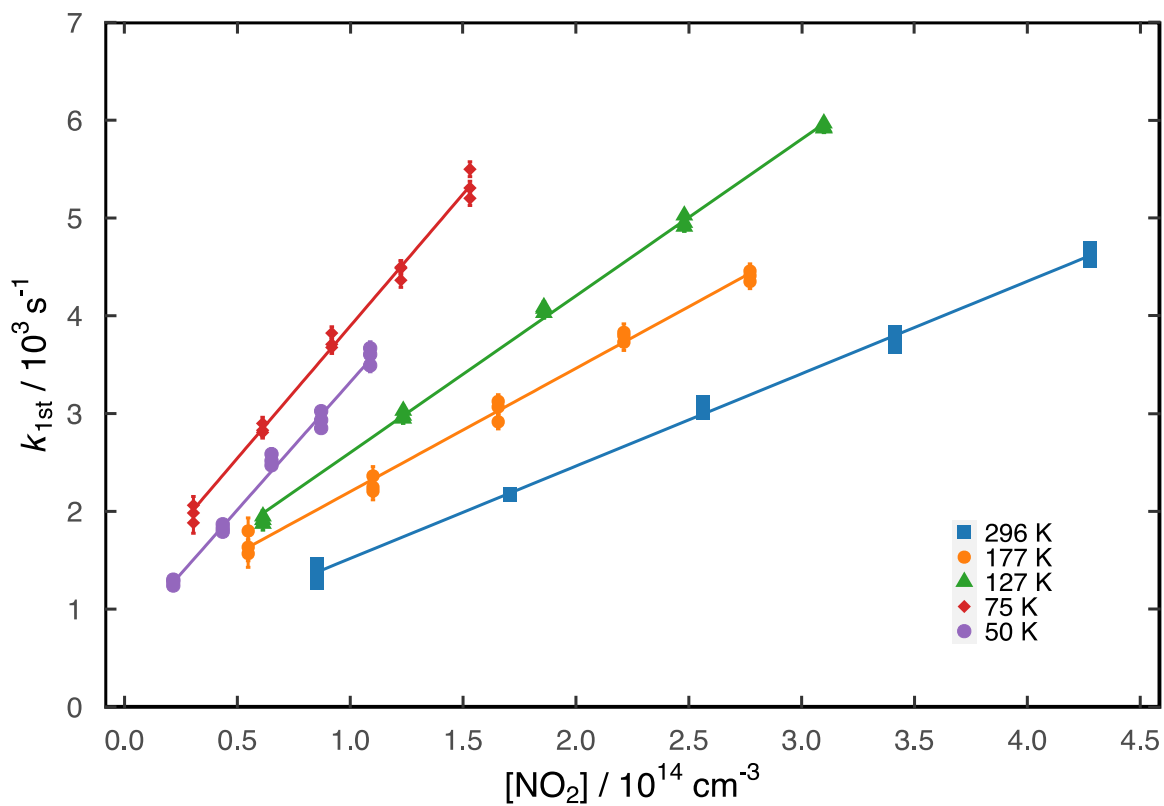

**Figure 5** Variation of the derived pseudo-first-order rate constant as a function of the NO$_2$ concentration for a range of temperatures. (Blue squares) 296 K; (orange circles) 177 K; (green triangles) 127 K; (red diamonds) 75 K; (purple circles) 50 K.

In a similar manner to the data analysis for the C + NO$_2$ reaction described in section 3.1, kinetic decays were recorded for 5 different values of the NO$_2$ concentration at any given temperature. The plots of the $k_{1st}$ values against the corresponding NO$_2$ concentration in Figure 5 show that $k_{1st}$ varies linearly as a function of the NO$_2$ concentration over the entire range and at all temperatures. Weighted linear least-squares fits to these datasets allowed us to derive second-order rate constants from the slopes.

The derived second-order rate constants are summarized in Table 3 alongside other relevant information and these data are displayed as a function of temperature alongside previous work in Figure 6.

**Table 3** Measured second-order rate constants for the O + NO$_2$ reaction

| T / K | $N^b$ | [NO]/10$^{14}$ cm$^{-3}$ | [NO$_2$]/10$^{13}$ cm$^{-3}$ | $k_{\mathrm{O+NO_2}}$/10$^{-11}$ cm$^3$ s$^{-1}$ | Carrier gas |
| --- | --- | --- | --- | --- | --- |
| 296 | 15 | 20.9 | 8.5 - 42.8 | 0.94 ± 0.1 | Ar |
| 177 ± 2 | 15 | 13.5 | 5.5 - 27.7 | 1.26 ± 0.13 | N$_2$ |
| 127 ± 2 | 15 | 15.1 | 6.1 - 30.9 | 1.60 ± 0.16 | Ar |
| 75 ± 2 | 15 | 7.5 | 3.1 – 15.3 | 2.70 ± 0.28 | Ar |
| 50 ± 1 | 15 | 10.7 | 2.2 - 10.9 | 2.61 ± 0.27 | Ar |

[a]Uncertainties on the calculated temperatures represent the statistical (1σ) errors obtained from Pitot tube measurements of the impact pressure. [b]Number of individual measurements.

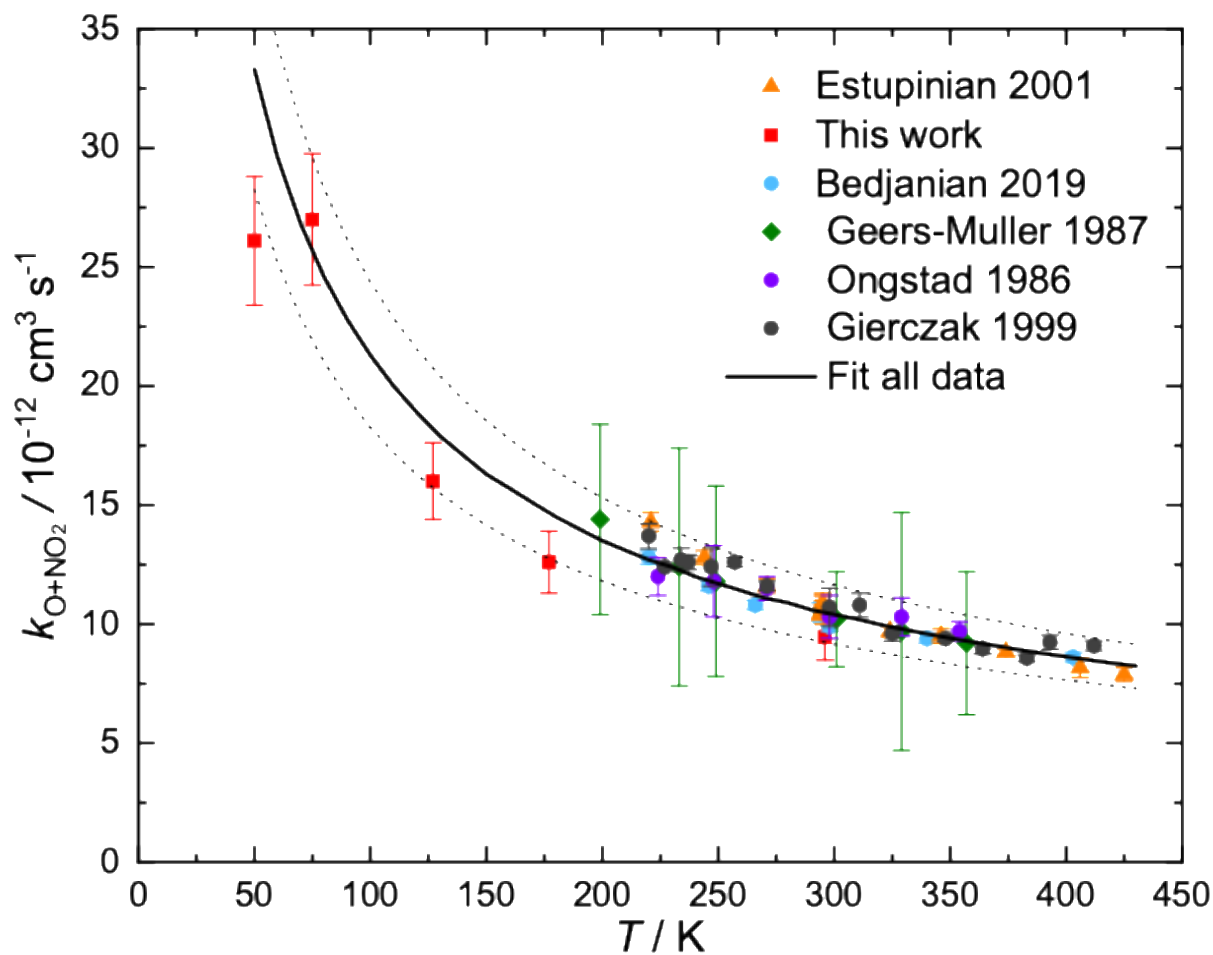

**Figure 6** Temperature dependence of the rate constants for the O + NO$_2$ reaction alongside selected previous studies. (Red squares) This work; (blue circles) Bedjanian & Kalyan[30]; (orange triangles) Estupiñán et al. [31]; (green diamonds) Geers-Muller & Stuhl[32]; (purple circles) Ongstad & Birks[33]; (grey circles) Gierczak et al. [34] The solid black line represents a fit to all of the data on this figure (so excluding the high temperature ( > 450 K) values of Bedjanian & Kalyan[31]) with an expression of the form $k(T) = \alpha(T/300)^\beta$. The dotted lines represent confidence intervals (68%) on the derived parameters.

It can be seen from Figure 6 that the room temperature rate constant derived in this work of $k_{O+NO_2}(296K)$ = (9.4 ± 1.0) × 10$^{-12}$ cm$^3$ s$^{-1}$ is in good agreement with the previous derived values of the earlier studies shown here in the range (9.9 – 10.9) × 10$^{-12}$ cm$^3$ s$^{-1}$, and with the currently recommended rate constant for this reaction of 10.4 × 10$^{-12}$ cm$^3$ s$^{-1}$. In common with the earlier studies conducted at lower temperature, the rate constants measured in the present work are also seen to increase as the temperature falls, reaching values of 2.6 and 2.7 × 10$^{-11}$ cm$^3$ s$^{-1}$ at 50 and 75 K respectively. As with the data for the C + NO$_2$ reaction, the lower measured rate constant value at 50 K is probably due to the increased difficulty of these measurements as explained earlier. Previous work by Gierczak et al.[34] has shown that this process presents little or no pressure dependence below 100 Torr, with no noticeable pressure dependence of their rate constants measured at 296 K or at 393 K.

A fit to all the experimental data[30-34] below 450 K of the form $k(T) = \alpha(T/300)^\beta$ can be used to represent the temperature dependence of the rate constants:

$$k(T) = (1.0 \pm 0.1) \times 10^{-11} (T/300)^{(-0.65 \pm 0.02)} \text{ cm}^3 \text{ s}^{-1} \qquad (E3)$$

If we use these parameters to extrapolate the fit down to 10 K, a rate constant of $k_{O+NO_2}(10\ K)$ = 9.5 × 10$^{-11}$ cm$^3$ s$^{-1}$ is obtained. It should be noted however that this extrapolation is outside of the valid temperature range (50-296 K) for these parameters.

**4 Astrochemical Model and Comparison with Observations**

To derive the abundances of NO$_2$ in interstellar environments, we used the gas-grain model Nautilus[35][36] in its three-phase form[37][38] to simulate the abundances of atoms and molecules in neutral and ionic form as a function of time, employing kida.uva.2014[39] as the basic reaction network. This network was updated recently for a better description of COMs on grains and in the gas-phase.[40-42] We use the following reactions involving the formation or destruction of NO$_2$ (main reactions only) to simulate NO$_2$ abundances in dense molecular clouds (in the 10-300K range):

| | | |
|---|---|---|
| NO + HO$_2$ → OH + NO$_2$ | $k_3$ = 6.0 × 10$^{-12}$ × (T/300)$^{-0.80}$ cm$^3$ s$^{-1}$ [43, 44] | (R2) |
| OH + HONO → H$_2$O + NO$_2$ | $k_4$ = 8.0 × 10$^{-12}$ × (T/300)$^{-0.60}$ cm$^3$ s$^{-1}$ [43, 45] | (R3) |
| H$^+$ + NO$_2$ → OH + NO$^+$ | $k_5$ = 1.6 × 10$^{-9}$ × (T/300)$^{-0.2}$ cm$^3$ s$^{-1}$ [46] | (R4) |
| H$_3^+$ + NO$_2$ → H$_2$ + OH + NO$^+$ | $k_6$ = 7.0 × 10$^{-10}$ × (T/300)$^{-0.2}$ cm$^3$ s$^{-1}$ [47] | (R5) |
| H + NO$_2$ → OH + NO | $k_7$ = 1.4 × 10$^{-10}$ cm$^3$ s$^{-1}$ [48, 49] | (R6) |
| N + NO$_2$ → NO + NO | $k_8$ = 1.1 × 10$^{-11}$ cm$^3$ s$^{-1}$ [13, 50] | (R7) |
| C + NO$_2$ → CO + NO | $k_9$ = 2.0 × 10$^{-10}$ × (T/300)$^{-0.11}$ cm$^3$ s$^{-1}$ (this work) | (R8) |
| O + NO$_2$ → NO + O$_2$ | $k_{10}$ = 1.0 × 10$^{-11}$ × (T/300)$^{-0.65}$ cm$^3$ s$^{-1}$ (this work) | (R9) |

The 800 individual species included in the network are involved in 9000 separate reactions. At the outset, elements are initially in atomic or ionic form (those elements with an ionization potential of less than 13.6 eV are considered as fully ionized), while the C/O elemental ratio is equal to 0.71. The initial simulation parameters are summarized in Table 4.

**Table 4** Astrochemical model parameters

| Element | Abundance[a] |
|---|---|
| H$_2$ | 0.5 |
| He | 0.09 |
| C$^+$ | 1.7 × 10$^{-4}$ |
| N | 6.2 × 10$^{-5}$ |
| O | 2.4 × 10$^{-4}$ |
| S$^+$ | 6.0 × 10$^{-7}$ |
| Fe$^+$ | 3.0 × 10$^{-9}$ |
| Cl$^+$ | 1.0 × 10$^{-9}$ |
| F | 6.7 × 10$^{-9}$ |

[a] Relative to total hydrogen (nH + 2nH$_2$)

The grain surface and the mantle are both chemically active for these simulations, while accretion and desorption are only allowed between the surface and the gas-phase. The dust-to-gas ratio (in terms of mass) is 0.01. The total H density (nH + 2nH$_2$) is set to 2.5 × 10$^4$ cm$^{-3}$, the temperature is 10 K, the cosmic ray ionization rate is 1.3 × 10$^{-17}$ s$^{-1}$, while the visual extinction is set to 10. A sticking probability of 1 is assumed for all neutral species while desorption can occur by thermal and non-thermal processes (cosmic rays, chemical desorption) including sputtering of ices by cosmic-ray collisions.[51] The surface reaction formalism and a more detailed description of the simulations can be found in Ruaud et al.[37]

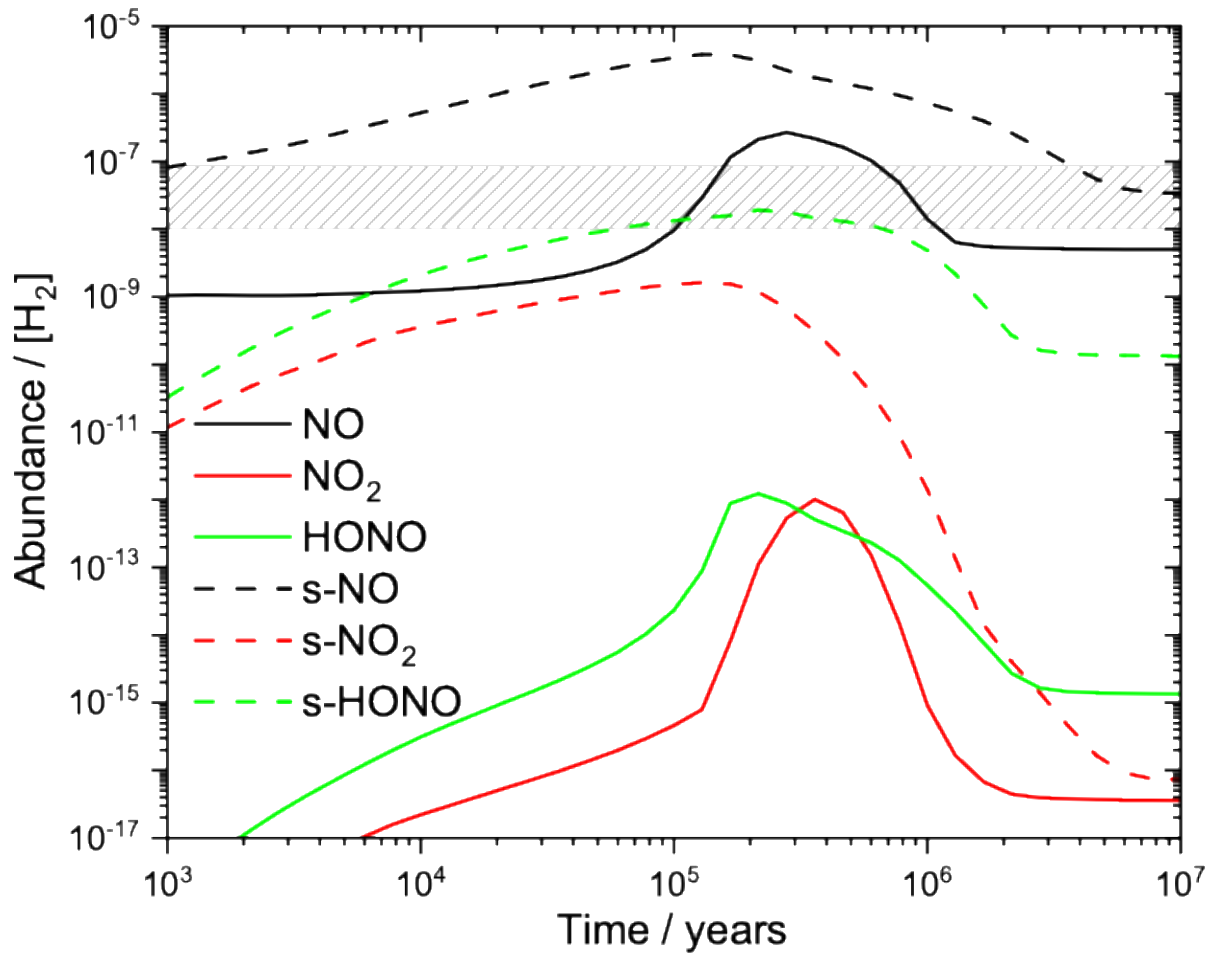

**Figure 7** Gas-grain astrochemical model results for the formation of NO, $NO_2$ and HONO as a function of time. Solid lines represent the gas-phase relative abundances (with respect to $H_2$), dashed lines represent the abundances present on dust grains. (Black lines) simulated NO abundance; (red lines) simulated $NO_2$ abundance; (green lines) simulated HONO abundance. The horizontal hatched rectangle represents the observed NO abundance in TMC-1 with an arbitrary error associated ($\pm\sqrt{3}$).

As shown in Figure 7, $NO_2$ is not a species that is characterized by high abundance levels in the gas-phase in the interstellar medium, with simulated abundances reaching a peak value of $10^{-12}$ with respect to $H_2$ around $4 \times 10^5$ years. In fact, there are no ionic reactions that directly or indirectly produce $NO_2$ [52] and the barrierless $NO + HO_2 \rightarrow OH + NO_2$ reaction (R2), which has a large rate constant at low temperature, [43, 44] involves very low fluxes due to the very low abundance of $HO_2$ detected only towards ρ-Ophiuci [53] and not in any dense molecular cloud. However, $NO_2$ is an important species on interstellar dust grains because it is formed at early to intermediate times by the accretion of O atoms on the grains to give s-O (s- means the species has accreted onto the grain) followed by the s-O + s-NO reaction, with s-NO itself being formed by the s-N + s-O reaction, these reactions being in competition with the more favorable N and O hydrogenation routes through reaction with s-H. Once formed on the grains, s-$NO_2$ is a source of s-HONO via the s-H + s-$NO_2$ reaction, HONO having been detected towards IRAS 16293-2422B. [7] Apart from the $NO + HO_2$ reaction, there is another minor route for producing $NO_2$ in the gas phase, the $OH + HONO \rightarrow H_2O + NO_2$ reaction (R3), inefficient due either to the very low abundance of HONO in the gas phase for dense molecular clouds, or the very low OH radical abundance around protostars where HONO is detected.

$NO_2$ appears to be a particularly difficult species to detect in the gas-phase interstellar medium. Firstly, it has a fairly small dipole moment (approximately 0.3 Debye). Secondly, it is highly reactive due to its radical nature, which severely limits its abundance (although other radical species, for example NO, can reach high abundance levels). In addition to the reactions (R8) and (R9) studied in this work (O + $NO_2$ and C + $NO_2$), and reaction (R5) with $H_3^+$, reaction (R6) between neutral atomic hydrogen and $NO_2$ is particularly important due to the large H-atom abundance (as high as $1 \times 10^{-4}$ with respect to $H_2$ with values close to $4 \times 10^{-5}$ even for evolved clouds > $10^6$ years). Considering the relatively high simulated abundance of $NO_2$ on interstellar ices, one potential target for the detection of $NO_2$ could be the soft shock zones described by Burkhardt et al. [54] In these environments, the grain mantle is thought to evaporate, allowing the species present on the grains to be injected into the gas phase without too much chemical modification. The relatively high $NO_2$ abundance on the grains could lead to a fairly high abundance in the gas-phase in these regions. However, to be detectable, the shock must be recent enough to avoid $NO_2$ being consumed by the reactions with atomic hydrogen and by depletion onto the dust grains.

**5 Conclusions**

This work describes an experimental investigation of the temperature dependent kinetics of the reactions of oxygen and carbon atoms in their ground triplet states with nitrogen dioxide. A supersonic flow reactor was used to perform these measurements over the 296-50 K temperature range. O and C atoms were generated directly in the supersonic flow through the pulsed laser photolysis of suitable precursor molecules introduced in low concentrations. C atoms were detected directly by vacuum ultraviolet laser induced fluorescence. In contrast, O atoms were followed indirectly through a chemiluminescence tracer method, employing the O + NO $\rightarrow$ NO$_2$* reaction to track the progress of the O + NO$_2$ reaction. The derived rate constants for the O + NO$_2$ reaction are in very good agreement with earlier work at higher temperature, displaying a negative temperature dependence over the entire range with a 3-fold increase at 50 K compared with the room temperature value. The measured rate constants for the C + NO$_2$ reaction are also seen to present a slight negative temperature dependence. The effects of these reaction on interstellar chemistry are tested by their inclusion in a gas-grain model of the dense interstellar medium. These results predict that the NO$_2$ abundance is likely to be too low for it to be detected in dense interstellar clouds. Nevertheless, as NO$_2$ is formed more efficiently on interstellar ices reaching relatively high abundance levels, it may be possible to detect NO$_2$ in certain warmer regions of interstellar space through sublimation of the dust grain mantles.

**Author information**

Corresponding Author

Kevin M. Hickson – Institut des Sciences Moléculaires ISM, CNRS UMR 5255, Univ. Bordeaux, F-33400 Talence, France; orcid.org/0000-0001-8317-2606;

Email: kevin.hickson@u-bordeaux.fr

Authors

Jean-Christophe Loison – Institut des Sciences Moléculaires ISM, CNRS UMR 5255, Univ. Bordeaux, F-33400 Talence, France

**Acknowledgements**

K.M.H. acknowledges support from the French program 'Physique et Chimie du Milieu Interstellaire'' (PCMI) of the CNRS/INSU with the INC/INP cofunded by the CEA and CNES as well as funding from the 'Program National de Planétologie'' (PNP) of the CNRS/INSU.